\documentclass[superscriptaddress,preprintnumbers,amsmath,amssymb,prb,twocolumn]{revtex4-1}
\usepackage[utf8]{inputenc}
\usepackage{enumitem}
\usepackage{graphicx,amsmath,color,multirow}

\def\w{\omega}

\def\ve{\varepsilon}

\def\<{\langle}
\def\>{\rangle}

\let\hide\iffalse
\usepackage{braket} 

\usepackage[caption=false]{subfig}

\begin{document}

\title{Electron-phonon couplings in polymorphous crystals}

\author{Marios Zacharias}
\email{zachariasmarios@gmail.com}
\affiliation{Univ Rennes, INSA Rennes, CNRS, Institut FOTON - UMR 6082, F-35000 Rennes, France}
\affiliation{Computation-based Science and Technology Research Center, The Cyprus Institute, Aglantzia 2121, Nicosia, Cyprus}
\author{George Volonakis}
\affiliation{Univ Rennes, ENSCR, INSA Rennes, CNRS, ISCR - UMR 6226, F-35000 Rennes, France}
\author{Laurent Pedesseau}
\affiliation{Univ Rennes, INSA Rennes, CNRS, Institut FOTON - UMR 6082, F-35000 Rennes, France}
\author{Claudine Katan}
\affiliation{Univ Rennes, ENSCR, INSA Rennes, CNRS, ISCR - UMR 6226, F-35000 Rennes, France}
\author{Feliciano Giustino}
\affiliation{ Oden Institute for Computational Engineering and Sciences, The University of Texas at Austin,
Austin, Texas 78712, USA
}%
\affiliation{Department of Physics, The University of Texas at Austin, Austin, Texas 78712, USA}
\author{Jacky Even}
\email{jacky.even@insa-rennes.fr}
\affiliation{Univ Rennes, INSA Rennes, CNRS, Institut FOTON - UMR 6082, F-35000 Rennes, France}

\date{\today}

\begin{abstract}


Positional polymorphism in solids refers to locally disordered unit cells that, on average, reproduce the high-symmetry 
structures observed in diffraction experiments. Standard theories of electron-phonon interactions fail to describe 
the temperature-dependent electronic structure of such polymorphous systems. Hybrid halide perovskites are a prime example, 
where configurational entropy from both polymorphism and molecular disorder plays a central role.
Here we generalize the special displacement method to polymorphous crystals, providing an efficient {\it ab initio} framework 
for electron-phonon couplings without resorting to molecular dynamics. We resolve long-standing 
discrepancies in hybrid halide perovskite physics, including temperature-dependent anharmonic phonons and band gaps.
Our approach provides a practical route to link local disorder, configurational entropy, and electron-phonon interactions, 
with applicability across diverse material classes, from optoelectronics and ferroelectrics to thermoelectrics.

\end{abstract}

\maketitle

Recently, the concept of correlated local disorder (positional polymorphism) 
has gained significant attention across a broad range of 
materials~\cite{Zacharias2025,Zhao2020,Wang2020_Z,Fabini2020,WangZhi2021,Goesten2022,Zacharias2023npj,Dirin2023,Sabisch2025,Wang2025}. 
In perovskites, local disorder 
has been observed experimentally since the 1960s~\cite{Comes1968}, and later confirmed through 
x-ray and neutron scattering as well as nuclear magnetic resonance measurements~\cite{Itoh1994,Mashiyama1998,Kiat2000,Worhatch2008,Beecher2016,Senn2016,Laurita2017,Sutton2018,Ferreira2020,Doherty2021,Dirin2023,Morana2023,Weadock2023,Yazdani2023,Xing2024,Sabisch2025,Dubajic2025}. 
A prominent class of such materials is hybrid halide perovskites~\cite{Mashiyama1998,Zhao2020}, 
which exhibit high efficiency and favorable optoelectronic properties, making them excellent candidates for energy 
applications such as photovoltaics and 
light-emitting devices~\cite{Kojima2009,Snaith2013,Lin2018,Sidhik2022,Metcalf2023,CorreaBaena2017,Mao2018}.
To realize the full potential of halide perovskites, a fundamental understanding of positional 
polymorphism, configurational entropy, and their interplay with the electronic, vibrational, 
and electron-phonon properties is essential. 

\begin{figure}[htb!]
 \begin{center}
\includegraphics[width=0.40\textwidth]{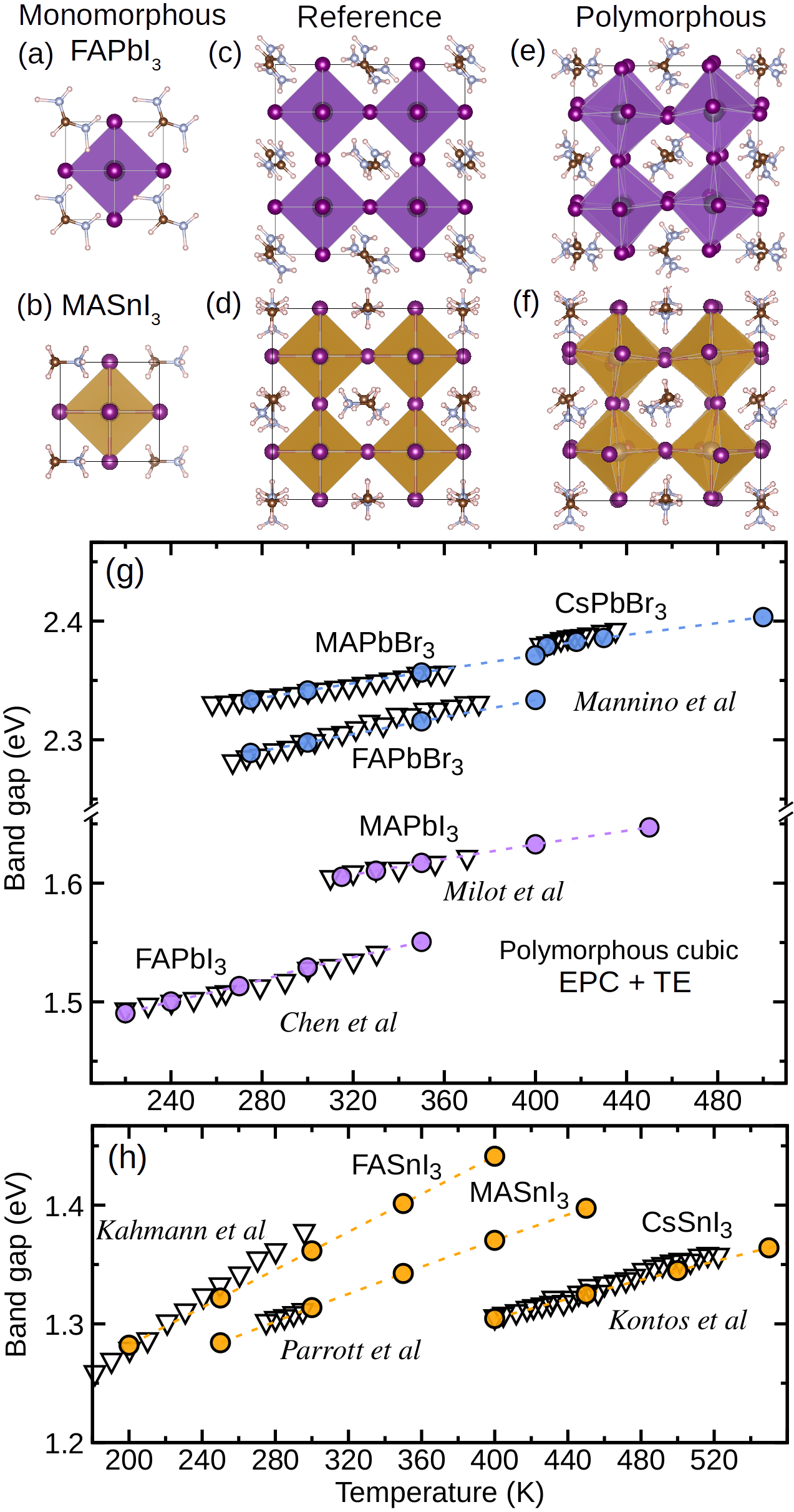}
 \end{center}  
\caption{(a-f) Ball-stick models of cubic FAPbI$_3$ and MASnI$_3$  representing 
 monomorphous (a,b), reference (c,d), and polymorphous (e,f) structures. Structures in (a,b) refer to 
unit cells (12 atoms), in (c,d,e,f) to 2$\times$2$\times$2 supercells. 
(g,h) Calculated temperature-dependent band gaps (coloured discs) of Pb (g) and Sn (h) cubic hybrid halide perovskites 
including electron-phonon coupling (EPC) and thermal expansion (TE) effects; EPC is incorporated via the ASDM using 
ZG polymorphous structures and TE contribution is evaluated by allowing lattice expansion according to the 
experimental expansion coefficients~\cite{Zacharias2026b}.
Experimental gaps (triangles) are from Refs.~[\onlinecite{Milot2015,Kontos2018,Parrott2016, Mannino2020, Chen2020,Kahmann2020}]. 
ASDM calculations refer to 4$\times$4$\times$4 supercells (768 atoms) and the DFT-PBEsol including 
SOC. Data are shifted close to PBE0 corrections (see Table~I of Ref.~[\onlinecite{Zacharias2026b}])
to match experiment. PBE0 corrections to electron-phonon 
renormalized band gaps of halide perovskites are negligible~\cite{Zacharias2023npj}.
\label{fig1} 
}
\end{figure}

Hybrid ABX$_3$ halide perovskites are composed of an organic molecule (A) 
and an inorganic network formed by metal cations (B) octahedrally coordinated by halide anions (X) [Figs.~\ref{fig1}(a,b)].
One central assumption in static density functional theory (DFT) calculations for 
tetragonal or cubic hybrid halide perovskites is the use of a periodic high-symmetry (monomorphous) 
metal-halide network~\cite{Even2013,Even2013b,Amat2014,Saidi2016,Whalley2017,Mao2018,Boubacar2019,Ponce2019,Ghaithan2020}.
The orientation of the molecule A is typically randomly chosen or set to towards an unrealistic static direction.
Critical issues related to the use of a high-symmetry inorganic network include 
(i) neglecting important corrections to the electronic structure~\cite{Zhao2020}, 
(ii) the breakdown of the standard harmonic approximation to describe lattice dynamics~\cite{Zacharias2023},  
(iii) neglecting energetically more stable
configurations and thus overlooking important structural features impacting chemical 
bonding and optoelectronic properties~\cite{Katan2018,Quarti2024,Garba2025}, 
and (iv) ignoring important modifications to electron-phonon coupling (EPC)~\cite{Zacharias2023npj}.

An approach for more accurate static DFT calculations using locally disordered (polymorphous) 
perovskites has been proposed~\cite{Zhao2020,WangZhi2021}. It has been shown that 
accounting for symmetry-breaking domains in tetragonal and cubic perovskites yields 
considerably improved agreement with measurements on pair distribution
functions, band gaps, and effective masses~\cite{Zhao2020,WangZhi2021,Zacharias2023npj}. 
Although molecular dynamics (MD)~\cite{Quarti2015,Quarti2016,Even2016,Wiktor2017,Carignano2017,Lahnsteiner2022,DiezCabanes2023,Weadock2023,Seidl2023,Xing2024,Dai2024} 
can capture local disorder through long equilibration times, a direct distinction between local 
disorder and thermal vibrations is inevitably missed. 

Recently, we have introduced a bottom-up first-principles approach~\cite{Zacharias2023npj}, showing that 
local disorder in {\it inorganic} halide perovskites (A=Cs) yields dynamically stable and strongly interacting phonons,
and affects profoundly anharmonic EPCs.
This polymorphous framework can be seen as a quasi-static approximation of the slow relaxation dynamics 
of the structure. In this Letter, we explore the more challenging and technologically 
relevant cases of hybrid halide perovskites and compare with their inorganic 
counterparts. Together with Ref.~[\onlinecite{Zacharias2026b}], we propose an efficient approach to 
evaluate phonons and temperature-dependent band gaps of polymorphous cubic APbBr$_3$, APbI$_3$, 
and ASnI$_3$, where A = methylammonium (MA) or formamidinium (FA). Our high-throughput calculations agree 
with experiments for all cases, as shown in Fig.~\ref{fig1}, opening the way for
effective predictions of anharmonic electron-phonon properties of polymorphous systems 
with configurational entropy. 



DFT calculations were performed using {\tt Quantum Espresso}~\cite{QE,QE_2} in conjunction 
with optimized norm-conserving Vanderbilt pseudopotentials~\cite{Haman_2013,vanSetten2018} and the Perdew-Burke-Ernzerhof exchange-correlation functional
revised for solids (PBEsol)~\cite{Perdew_2008}; full computational details are available in Ref.~[\onlinecite{Zacharias2026b}]. 
In Figs.~\ref{fig1}(a)-(f), we show the geometries of the cubic FAPbI$_3$ and MASnI$_3$. 
We consider three cases: (i) the monomorphous structure represented by a minimal unit cell [i.e. the fomula unit (f.u.)]
where the atoms of the inorganic network reside at their high-symmetry positions and the FA or MA molecules point towards 
a single orientation [Figs.~\ref{fig1}(a) and (b)], (ii) the reference structure represented by a supercell
where the atoms of the inorganic network are fixed at their high-symmetry positions and the FA or MA molecules are 
relaxed starting from random orientations [Figs.~\ref{fig1}(c) and (d)], (iii) the polymorphous structure represented by a supercell
where the atoms of the inorganic network and FA or MA molecules are relaxed starting from the reference structure 
[Figs.~\ref{fig1}(e) and (f)].
During relaxations, the lattice constants are fixed at their experimental values~\cite{Zacharias2026b}.
We consider reference configurations since changes in the electronic structure and phonons 
are sensitive to the net dipole moments induced by single cation orientations 
and to ensure a consistent comparison with polymorphous structures. 
In Cs-based compounds, the reference is identical to the monomorphous structure so no special treatment is required.
Additionally, to assess the electronic structure and thermal effects variability 
of the polymorphous networks originating from configurational entropy~\cite{Quarti2015,Motta2015}, 
we generate 10 distinct polymorphous structures and take Boltzmann-weighted averages 
relying on the system's free energy~\cite{Zacharias2026b}. 
Convergence tests of the band gap as a function of the number of polymorphous configurations 
are provided in Ref.~[\onlinecite{Zacharias2026b}].

To compute thermal band gap renormalization, we consider both EPC and lattice thermal expansion (TE) so that 
the band gap temperature coefficient is given by~\cite{Villegas2016}:
\begin{equation} \label{eq.1}
 {d E_{\rm g}}/{ dT} = {d E_{\rm g}}/{ dT}|_{\rm EPC} + {d E_{\rm g}}/{ dT}|_{\rm TE}.
\end{equation}
We use temperature-dependent phonons computed within the self-consistent phonon theory as implemented 
in the anharmonic special displacement method~\cite{Zacharias2023} (ASDM).
This requires computing interatomic force constants iteratively of the corresponding thermally
displaced reference structures until convergence in the phonons and system's free energy~\cite{Zacharias2026b}. 
The computed dynamically stable phonons allow us to include phonon anharmonicity in nonperturbative calculations of 
EPC via the special displacement method~\cite{Zacharias2020,Lee2023,Zacharias2023npj}. 
That is we apply special Zacharias-Giustino (ZG) displacements~\cite{Zacharias2026b} starting  
from the polymorphous structures. To account for TE, we expand the lattice based on the experimental expansion 
coefficients~\cite{Kawamura2002,Mashiyama2003,Stoumpos2013,Rakita2015,Whitfield2016,Dang2016,Fabini2016,Schueller2017,Kontos2018,Marronnier2018,Keshavarz2019,Handa2020,He2021}, as decribed in Ref.~[\onlinecite{Zacharias2026b}]. 

In Figs.~\ref{fig1}(g,h),  we compare our calculations for temperature-dependent band gaps of polymorphous cubic 
MAPbBr$_3$, MAPbI$_3$, FAPbBr$_3$, FAPbI$_3$, FASnI$_3$, and MASnI$_3$ with 
experiments~\cite{Milot2015,Kontos2018,Parrott2016, Mannino2020, Chen2020,Kahmann2020}. We
include data for CsPbBr$_3$ and CsSnI$_3$ for completeness~\cite{Zacharias2026b}. 
Calculations and experiments agree for all compounds, showcasing the broad applicability 
of our approach in describing anharmonic electron-phonon renormalized band structures. 
We note that an empirical shift
is applied only to align the absolute band gap values with experiment.
Quantitative agreement of temperature-dependent trends is evidenced by our 
calculated $dE_{\rm g} / dT$ reported in Table~\ref{table.1}, in which  
we further provide data for CsPbI$_3$~\cite{Zacharias2026b}. We also show the EPC contribution to the gap 
renormalization for all materials, varying between 27\,--\,97\%. 
For the majority of compounds, EPC {\it plays a dominant role} in band gap renormalization, 
challenging the prevailing view that TE and EPC 
contribute equally in lead halide perovskites~\cite{FranciscoLpez2019,Rubino2020}.
A clear trend is observed in the variation of the EPC contribution across all compounds, with its
percentage increasing as the A-site cation changes from FA to MA and then to Cs.
For example, in FAPbBr$_3$ and FASnI$_3$, EPC contributes only 27\% and 34\%, respectively. 
This lower contribution is not surprising, as their volumetric TE coefficients are among 
the highest reported for halide perovskites~\cite{Schueller2017}.
Sn-based compounds exhibit a larger $dE_{\rm g} / dT$, which we attribute to positional polymorphism enhancing 
band gap renormalization via TE. This effect is driven by the stronger lone pair activity~\cite{Balvanz2024} of Sn
that promotes Sn-I-Sn bending under TE~\cite{Zacharias2026b}.
We note that small deviations from experiment arise from approximations in our 
framework (e.g. constant-temperature phonons or neglect of non-adiabatic and many-body 
effects~\cite{Antonius2014,Miglio2020}), and from limited experimental 
data in some cases such as CsPbI$_3$.

\begin{table}[t!]
\caption{Band gap temperature coefficient (${d E_{\rm g}}/{ dT}$) of cubic halide perovskites. r- and p-
stand for reference and polymorphous. Experimental values are extracted by linear fits~\cite{Zacharias2026b}. }
\setlength{\arrayrulewidth}{0.6pt} 
\begin{tabular}{ l c c c c}
\hline \hline
       & ${d E_{\rm g}}/ { dT}\big|_{\rm expt.}$ & \,\,$ {d E_{\rm g}}/ { dT}$ \,\,  &  $ {d E_{\rm g}} / { dT}\big|_{\rm EPC}$  & \,\, EPC \,\,   \\
 & \multicolumn{3}{ c}{ ($10^{-4}$ eV/K)}  & \,\, Pct. \,\,  \\
\hline 
r-MAPbBr$_3$ & 2.8 & 5.8 & 4.0  & 71\%   \\  
p-FAPbBr$_3$ & 4.9 & 3.6 & 1.2  & 34\%   \\ 
p-MAPbBr$_3$ & 2.8 & 2.9 & 2.0  & 70\%   \\ 
p-CsPbBr$_3$ & 3.4 & 2.6 & 2.4  & 94\%   \\
p-FAPbI$_3$  & 4.3  & 4.6 &  3.2 & 70\%   \\  
p-MAPbI$_3$  & 2.6 & 3.1 &  2.7 & 88\%    \\ 
p-CsPbI$_3$  & 8.4  & 1.2  & 1.2  & 97\%    \\  
p-FASnI$_3$  & 10.6  & 8.0 &  2.2 & 27\%   \\  
p-MASnI$_3$  & 4.5  & 5.7 &  3.5 & 61\%   \\  
p-CsSnI$_3$  & 4.7  & 3.9 &  3.3 & 84\%   \\  
\hline
\hline
\end{tabular}
\label{table.1}
\end{table}

\begin{figure}[b!]
 \begin{center}
\includegraphics[width=0.44\textwidth]{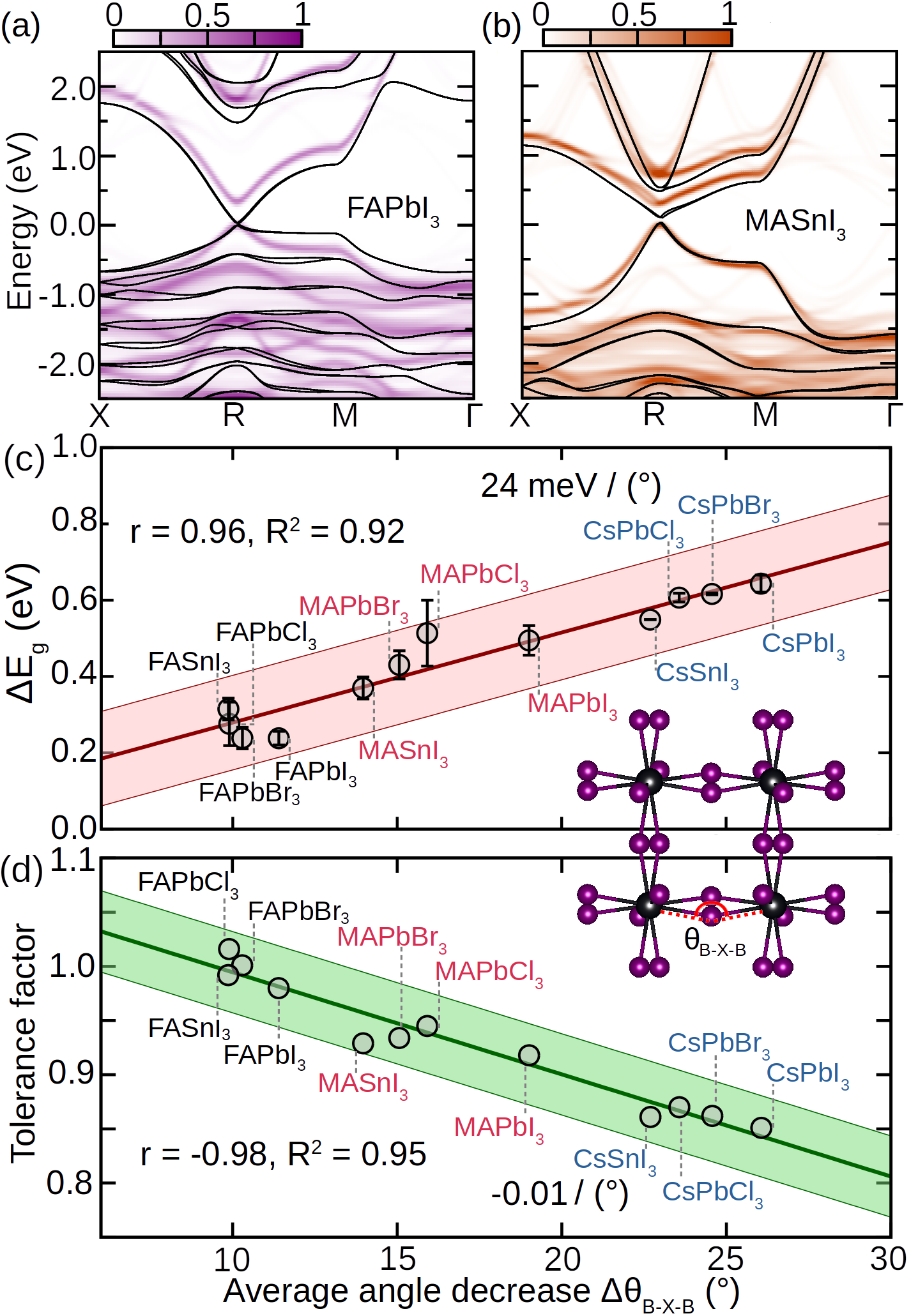}
 \end{center} 
\caption{ 
(a,b) Electron spectral functions (color maps) of the polymorphous cubic FAPbI$_3$ (a) and MASnI$_3$ (b). 
Black dispersions represent the band structure of the monomorphous networks. 
Calculations are at the PBEsol+SOC level and thus underestimate the band gap. Combining hybrid functionals
with SOC and polymorphous structures reproduce experiments well~\cite{Zacharias2026b,Garba2025}.
(c,d) Average band gap increase due to positional polymorphism $\Delta E_g$ (c) and tolerance factor $t$ (d)
versus 
the average metal-halide-metal angle decrease $\Delta \theta_{\rm B-X-B}$ calculated for 10 distinct polymorphous configurations. 
Solid lines are fits obtained by linear regression and shaded regions represent three standard deviations
on either side of the lines. 
The error bars represent the standard deviation across 10 configurations used for each material.
\label{fig2} 
}
\end{figure}

Now we analyze the effect of local disorder in the electronic structure of hybrid halide perovskites 
without thermal effects. In Figs.~\ref{fig2}(a) and (b), we compare the electron spectral 
functions (color maps) obtained for the polymorphous
cubic FAPbI$_3$ and MASnI$_3$ with the band structures (black lines) calculated 
for their monomorphous networks. Spectral functions were calculated using band unfolding~\cite{Popescu2012} 
including spin orbit coupling (SOC) as implemented in {\tt bands\_unfold.x} of the {\tt EPW/ZG} module~\cite{Zacharias2020,Lee2023}. 
Local disorder leads to significant 
changes in the electronic structure, including enhanced band broadenings and gap openings. Particularly, our calculations 
for the polymorphous cubic FAPbI$_3$ yield an average hole and electron effective mass 
enhancements~\cite{Giustino2017,Zacharias2026b} of $\lambda_{\rm h} = 2.3$  and $\lambda_{\rm e} = 2.0$,
and a gap opening of $\Delta E_{\rm g} = 0.24$~eV compared to the reference structure. 
The latter value compares well with $\Delta E_{\rm g} = 0.29$~eV with respect to the monomorphous structure,
 reported in Ref.~[\onlinecite{Zhao2020}]. The corresponding results for polymorphous MASnI$_3$
are $\Delta E_{\rm g} = 0.37$~eV,  $\lambda_{\rm h} = 1.2$, and $\lambda_{\rm e} = 3.0$.

Figure~\ref{fig2}(c) show the dependence of the band gap increase due 
to positional polymorphism $\Delta E_{\rm g}$ in hybrid halide perovskites 
on the 
average bond angle decrease $\Delta \theta_{\rm B-X-B}$. 
For completeness, we also include data for Cs- and Cl-based halide perovskites from Ref.~[\onlinecite{Zacharias2026b}].
The measure~\cite{Baur1974,Filip2014} $\Delta \theta_{\rm B-X-B}$ is strongly correlated with bond length changes~\cite{Zacharias2026b}
and thereby is chosen as the main descriptor for quantifying local disorder with respect to the high-symmetry inorganic network.
Our calculations and analysis yield $\Delta E_{\rm g}$ in the range 
198--616~meV and a slope 
of 22~meV/($^\circ$), revealing a strong correlation between $\Delta E_{\rm g}$ and bond angles change 
with a Pearson correlation of $r=0.96$.
This is expected, as deviations in bond angles from their ideal values  
lead to a reduction in the antibonding coupling between metal and halogen orbitals~\cite{Knutson2005,Filip2014,Dar2016,Meloni2016,Zacharias2026b}. 
The high $R^2$ of 0.92 show that this distortion measure can effectively explain the gap opening due to positional polymorphism.
Furthermore, the band gap variability, computed for 10 configurations and shown as error bars [Fig.~\ref{fig2}(c)], is larger for 
MA-based compounds due to the larger dipole moments of the MA compared to the FA cations~\cite{Carignano2017,Maheshwari2019}.
Cl compounds, e.g. MAPbCl$_3$, exhibit the largest band gap variability under local disorder because 
the smaller, more electronegative Cl ions create stronger Pb-Cl bonds, making the band structure 
highly sensitive to structural distortions.

Compared to their reference structures, polymorphous cubic FA, MA, and Cs-based compounds 
exhibit total energy lowerings of 50--73 meV/f.u., 53--93 meV/f.u., and 61--121 meV/f.u., respectively~\cite{Zacharias2026b}. 
This trend highlights the role of A-cation size in stabilizing locally disordered phases, with smaller cations 
facilitating structural distortions~\cite{Kieslich2014,Li2022}.
This is further illustrated in Figs.~\ref{fig2}(c) and (d), where bond angle variations 
depend strongly on the cations and the tolerance factors.  
We evaluate the tolerance factors ($t$) using 
ionic radii of 1.19, 1.15, 2.2, 1.96, 1.81, 1.88, 2.2, and 2.5~\AA\,
for Pb$^{+2}$, Sn$^{+2}$, I$^{-}$, Br$^{-}$, Cl$^{-}$, Cs$^{+}$, MA$^{+}$, and FA$^{+}$~\cite{Shannon1976,Kieslich2015}, respectively.
Our analysis suggests that smaller tolerance factors allow for greater local disorder, and thus larger band gap openings.
For instance, the relatively low tolerance factors of Cs-based halide perovskites ($t = 0.85$–$0.87$) 
facilitate octahedral distortion due to the absence of larger organic A-site cations~\cite{Kieslich2014,Filip2018,Li2022}. This structural flexibility promotes a higher 
local disorder, which leads to significant band gap renormalizations exceeding 0.55 eV.
Similarly, MA-based compounds exhibit a higher degree of polymorphism, smaller $t$, and larger $\Delta E_{\rm g}$
than FA-based compounds, attributable to the smaller size of MA.

\begin{figure}[b!]
 \begin{center}
\includegraphics[width=0.495\textwidth]{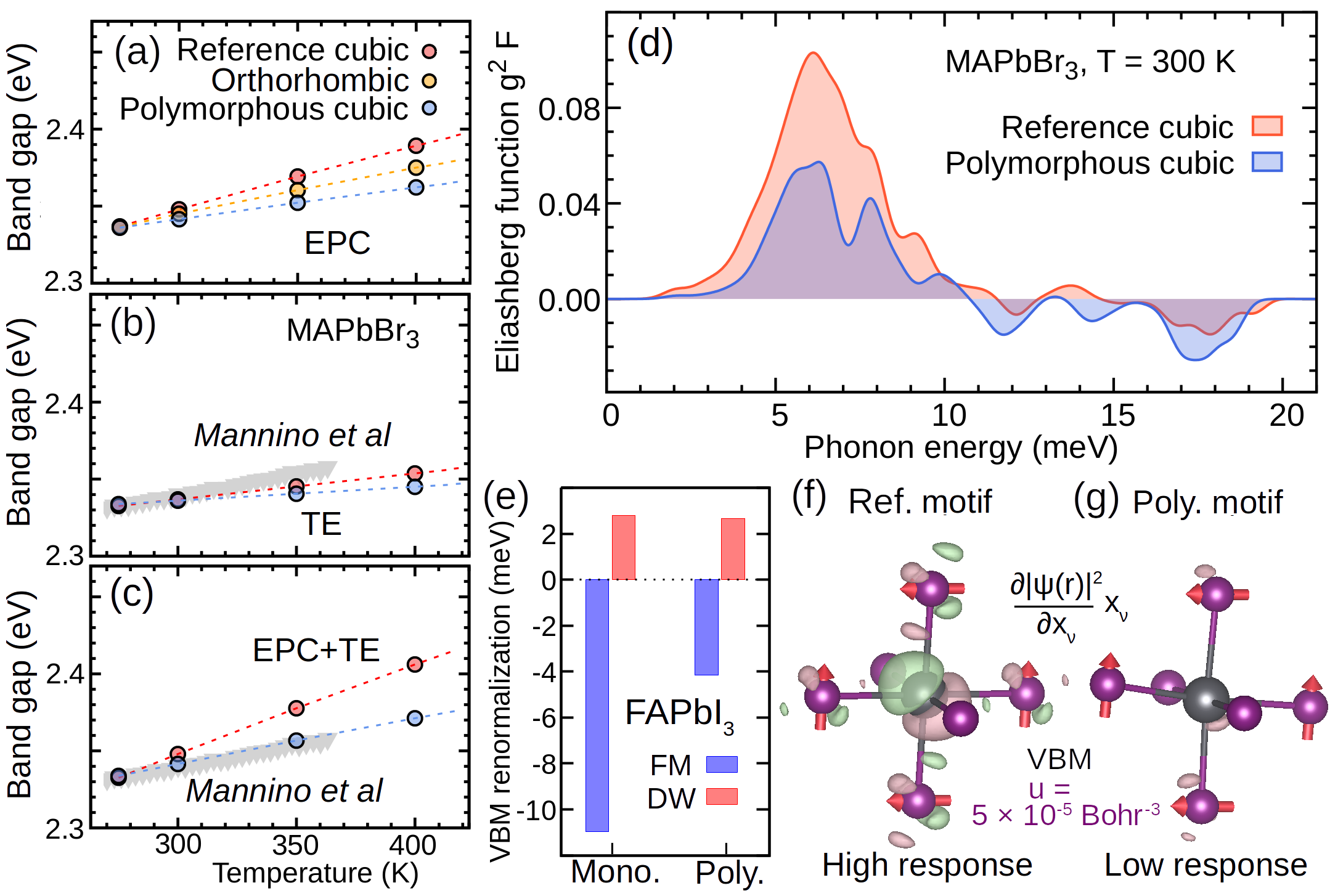}
 \end{center}  
\caption{
(a,b,c) Temperature-dependent band gap of cubic MAPbBr$_3$ calculated considering (a) 
electron-phonon coupling (EPC), (b) lattice thermal expansion (TE), and (c) both effects combined. 
Calculated data are shifted by 1.38 eV (orthorhombic), 1.79 eV (monomorphous cubic) 
and 1.44 eV (polymorphous cubic) to match the experiment~\cite{Mannino2020} (grey triangles) at $T=275$~K. 
(d) Eliashberg function showing the one-phonon energy-resolved band gap renormalization of MAPbBr$_3$ at $T=300$~K. 
(e) Fan-Migdal (FM) and Debye-Waller (DW) self-energy 
corrections to the VBM of the reference and polymorphous cubic FAPbI$_3$  due to a bending mode.
(f,g) Linear variations in the charge density at the VBM 
obtained for a single motif of the reference (f) and polymorphous (g) networks. Arrows representing 
the bending mode are scaled by 50. The isosurface value $u$ is indicated. 
\label{fig3} 
}
\end{figure}

At this point, we emphasize that local disorder in hybrid halide perovskites appears to be a key factor 
in determining the band gap order between FA- and MA-based compounds. For example, as shown in Fig.~\ref{fig1}(g), 
MA-based cubic lead halide perovskites have a larger band gap compared to their FA-based counterparts 
within the same temperature range.  The difference can be attributed to the band gap widenings 
reported in Fig.~\ref{fig2}(c). In the case of tin 
halide perovskites, the situation is reversed, a trend that agrees with experiments, as shown in Fig.~\ref{fig1}(h).
This observation is explained, e.g., by the larger TE coefficient of FASnI$_3$~\cite{Schueller2017}, 
together with the greater propensity for lone pair expression in sterically constrained FA-based compounds, 
which enhance thermal-induced band gap renormalization.  

We next elucidate the mechanisms determining the band gap temperature coefficient of halide perovskites.
Figure~\ref{fig3}(a) shows the electron-phonon renormalized band gap of MAPbBr$_3$ as a function of temperature calculated for 
the ZG reference (red) and ZG polymorphous (blue) cubic structures, both generated using stable anharmonic phonons. Local disorder 
significantly affects EPC, leading to a slower band gap variation with temperature. 
On the same plot, we include calculations for the orthorhombic structure (orange) using 
harmonic phonons. Attempting to mimic positional polymorphism at high temperatures 
using the orthorhombic phase yields improved results but does not fully
accurately capture the effect of EPC. Notably, polymorphism also affects the band gap widening due to TE, as shown 
in Fig.~\ref{fig3}(b);
a point that has been overlooked in previous calculations~\cite{Foley2015,Saidi2016,FranciscoLpez2019,Ning2022}. 
Accounting for local disorder in our calculations reduces the band gap opening due to TE 
by 50\% and the slope $dE_{\rm g} / dT|_{\rm TE}$ alone underestimates significantly the experimental data~\cite{Mannino2020}. 

In Fig.~\ref{fig3}(c), we show that combining EPC and lattice expansion in our calculations for the 
polymorphous cubic MAPbBr$_3$, the total $dE_{\rm g} / dT$ yields 
excellent agreement with experiment. In contrast, $dE_{\rm g} / dT$ is overestimated 
by a factor of two when the ZG reference network is used. 
A quantitative comparison of $dE_{\rm g} / dT$ alongside the percentage contribution of 
EPC for reference and polymorphous MAPbBr$_3$ is presented in Table~\ref{table.1}.
 
Figure~\ref{fig3}(d) shows the one-phonon energy-resolved contribution to the band gap renormalization 
of MAPbBr$_3$ at $T=300$~K calculated using the Eliashberg function:
 $ g^2F_n(\w,T) = \sum_\nu \Delta \ve_{n,\nu} (T) \delta(\hbar \w - \hbar \w_\nu).$
Here, $\hbar \w_\nu$ is the phonon energy of mode $\nu$, $\Delta \ve_{n,\nu}(T) = \frac{1}{2} \frac{\partial^2 \ve_n}{\partial x_\nu^2} \sigma^2_{\nu,T} $ is the mode-resolved energy renormalization of a state $\psi_n$ up to second order, 
and  $\sigma^2_{\nu,T}$ is the associated mean-square displacement~\cite{Zacharias2016}. 
$\Delta \ve_{n,\nu}(T)$ is calculated by finite differences using
single mode displaced configurations generated by {\tt ZG.x}~\cite{Lee2023}. Our results for the 
reference and polymorphous structures reveal that low energy optical vibrations in the window 
2--10~meV and 4--10~meV, respectively, dominate the phonon-induced band gap widening, in agreement with 
interpretations of previous works~\cite{Zacharias2023npj,Yazdani2023,Zhu2024, Biswas2024}. 
The modes associated with this energy window have a bending or rocking character associated with 
 octahedral tilting~\cite{PrezOsorio2015}. Higher frequency modes 
are mostly associated to B-X stretching or molecular librations. 
Positional polymorphism strongly suppress the positive EPC contribution in the window 2--10~meV,
while enhancing the negative contributions in the window 10--20~meV, resulting 
in significant reduction in the overall gap opening. 

To elucidate the origin of this reduction, 
we investigate the band gap renormalization due to a mode 
with bending character. 
If we consider the Kohn-Sham Hamiltonian $H_{\rm KS}$, express $\ve_{n} = \braket{\psi_n | H_{\rm KS} | \psi_n}$, 
take the second derivative with $x_\nu$, 
and apply the Hellmann-Feynman theorem twice~\cite{Ponce2014} we obtain:
\begin{eqnarray} \label{eq.1}
\hspace{-0.4cm} \frac{\partial^2 \ve_n}{\partial x_\nu^2} =
\int d{\bf r} \frac{\partial V_{\rm KS}}{\partial x_\nu} \frac{\partial |\psi_n({\bf r})|^2} {\partial x_\nu}
+   \int d{\bf r} \frac{\partial^2 V_{\rm KS}}{\partial x_\nu^2} |\psi_n({\bf r})|^2. 
\end{eqnarray}
Here, $V_{\rm KS}$ is the Kohn-Sham potential and the integration is taken over all electron coordinates ${\bf r}$. 
To arrive at this expression we also neglected the non-local contribution to $V_{\rm KS}$~\cite{Starace1971}.
The first and second terms on the right hand side of Eq.~\eqref{eq.1} are the Fan-Migdal (FM) and Debye-Waller (DW)
EPC coefficients~\cite{Giustino2017}.


Figure~\ref{fig3}(e) shows the FM and DW self-energy corrections to the valence band maximum (VBM)
induced by nuclei vibrations of amplitude 0.02 \AA~along a bending mode. We present results 
for the reference and polymorphous networks of FAPbI$_3$. In both cases, the negative FM dominates 
over the positive DW correction leading to an overall energy decrease of the VBM of 8.2~meV and 1.5~meV for the 
reference and polymorphous networks, respectively. This contributes to the band gap opening in both cases. The relatively 
large difference in the VBM energy decrease is associated to the strong reduction of the FM term (62\%) when we 
employ a polymorphous network. This result can be explained by inspecting the changes of the electronic charge distribution 
at the VBM in response to the movement of the nuclei along the bending mode 
(i.e. $\partial |\psi({\bf r})|^2 / \partial x_\nu$), shown in Figs.~\ref{fig3}(f) and (g). 
The charge density in the reference structure exhibits a higher response to the nuclear vibrations 
than that in the polymorphous structure, leading essentially to a much larger self-energy correction. 
The reduced response in the polymorphous structure arises from the deviation of Pb-I-Pb bond angles 
from 180$^\circ$, which disrupts orbital overlap and localizes the electronic states, thereby weakening 
the electron-phonon coupling. On the other hand, the DW corrections to the VBM of the 
reference and polymorphous structures are nearly the same [Fig.~\ref{fig3}(e)], as the
charge density $|\psi({\bf r})|^2$ remains the same and concentrated around the I atoms.  
A similar behavior is observed for the conduction band minimum (CBM), where the DW terms 
remain similar while the positive FM term, leading to a CBM energy increase, is reduced by 71\% due to local disorder. 

In conclusion, we generalize the ASDM for the simultaneous treatment of 
anharmonicity  and configurational entropy in both the inorganic and organic sublattices, as well 
as electron-phonon coupling, without long path-integral or classical MD runs. 
We clarify the origin of the gradual band gap blueshift of halide perovskites with temperature, 
attributing it to the low response of electrons to nuclear vibrations induced by positional polymorphism.
We elucidate the influence of the A-site cation on local disorder, band gap,  
and electron-phonon renormalization, establishing rules for materials engineering.
In the future, it will be interesting to investigate temperature-dependent 
properties of emerging hybrid materials 
with optoelectronic, ferroelectric, and thermoelectric applications~\cite{Haque2020,Sidhik2022,Zhang2022,Dirin2023,Metcalf2023,Jiang2025}.
Our approach forms the basis for electron-phonon calculations in polymorphous 
 solids, including phonon-assisted absorption~\cite{Zacharias2015,Noffsinger2012,Zacharias2016}, 
mobilities~\cite{Herz2017,Schlipf2108,Ponce2019,Cucco2024}, polarons~\cite{Franchini2021,RendeCotret2022,Jon2022,Zhang2023,Seiler2023,Biswas2024}, and 
exciton-phonon interactions~\cite{Alvertis2023,Zhenbang2024}. 

Data leading to the results of this study is available in the NOMAD repository~\cite{Zacharias_NOMAD_2025}.

\vspace*{0.3cm}

\acknowledgments
\vspace*{-0.2cm}
This research was funded by the European Union (project ULTRA-2DPK / HORIZON-MSCA-2022-PF-01 /
Grant Agreement No. 101106654). Views and opinions expressed are however those of the authors only and do not necessarily 
reflect those of the European Union or the European Commission. Neither the European Union nor the 
granting authority can be held responsible for them.
We thank D. R. Ceratti for graciously providing data on 
temperature-dependent band gaps of Br-based hybrid halide perovskite single crystals.
J.E. acknowledges financial support from the Institut Universitaire de France.
F.G. was supported by the Robert A. Welch Foundation under Award No. F-2139-20230405 and by the National Science Foundation under DMREF Grant No. 2119555.
G.V. acknowledges funding from the ANR through the CPJ program and the SURFIN project (ANR-23-CE09-0001), 
the ALSATIAN project (ANR-23-CE50-0030), and the ANR under the France 2030 programme, MINOTAURE project (ANR-22-PETA-0015).
We acknowledge that the results of this research have been achieved using computational resources from the EuroHPC Joint Undertaking 
and supercomputer LUMI [https://lumi-supercomputer.eu/], hosted by CSC (Finland) and the LUMI consortium through a EuroHPC Extreme Scale Access call.

\bibliography{references}{}

\end{document}